# Direction-Dependent Stability of skyrmion lattice in helimagnets induced by Exchange Anisotropy

Yangfan Hu[*]

Sino-French Institute of Nuclear Engineering and Technology, Sun

Yat-Sen University, 510275 GZ, China

Exchange anisotropy provides a direction dependent mechanism for the stability of the skyrmion lattice phase in noncentrosymmetric bulk chiral magnets. Based on the Fourier representation of the skyrmion lattice, we explain the direction dependence of the temperature-magnetic field phase diagram for bulk MnSi through a phenomenological mean-field model incorporating exchange anisotropy. Through quantitative comparison with experimental results, we clarify that the stability of the skyrmion lattice phase in bulk MnSi is determined by a combined effect of negative exchange anisotropy and thermal fluctuation. The effect of exchange anisotropy and the order of Fourier representation on the equilibrium properties of the skyrmion lattice is discussed in detail.

**I. Introduction**

Skyrmions are topologically protected solitons which can appear in magnetic systems due to the antisymmetric Dzyaloshinskii-Moriya (DM) interactions [1, 2]. Magnetic skyrmions are attractive for their topological Hall effect [3, 4], their sensitivity to electric current[5, 6], their stability in nanomaterials[7-11], and their tunability to various kinds of field[12-16]. In bulk chiral magnets[14, 17-19], and corresponding thin films[20-24], 2D magnetic skyrmions are found to spontaneously form crystalline state referred to as the skyrmion lattice phase. Theoretically, the skyrmion lattice phase is metastable compared with the conical phase when analyzed within a mean-field phenomenological theory[25]. It is shown that when effects of thermal fluctuation are considered, the skyrmion lattice phase becomes stable in certain area of the temperature-magnetic field phase diagram[17]. The effect of thermal fluctuation on the stability of the skyrmion lattice phase is independent of the direction of applied magnetic field, for which the distinction between the phase diagrams of MnSi detected when the magnetic field is applied in (100), (110), and (111) [26, 27] is not understood. A possible effect which provides a direction dependent stability of the skyrmion lattice phase is the cubic exchange anisotropy[2, 28].

Based on the Fourier representation of the skyrmion lattice phase[29], we show that the experimentally observed phase diagram of bulk MnSi when the magnetic field is applied in (100) [17] can be quantitatively obtained within a phenomenological mean-field theory. We find that by introducing a weak exchange anisotropy with negative coefficient in the free energy functional, the pocket-like area of stable skyrmion lattice phase in bulk MnSi can be reproduced not only qualitatively but also quantitatively.

**II. Formulation**

We use the following free energy density functional to study magnetic skyrmions in cubic helimagnets[1, 17, 28, 30]

$$w(\mathbf{M}) = \sum_{i=1}^{3} A \left(\frac{\partial \mathbf{M}}{\partial x_i}\right)^2 + D\mathbf{M} \cdot (\boldsymbol{\nabla} \times \mathbf{M}) - \mathbf{B} \cdot \mathbf{M} + \alpha(T - T_0)\mathbf{M}^2 + \beta \mathbf{M}^4 + \sum_{i=1}^{3} A_e \left(\frac{\partial M_i}{\partial x_i}\right)^2, \quad (1)$$

where $\mathbf{M} = [M_1 \quad M_2 \quad M_3]^T$ denotes the magnetization vector. The six terms on the right-hand side of eq. (1) describe respectively the exchange energy density, the Dzyaloshinskii-Moriya (DM) coupling, the Zeeman energy density with applied magnetic field **B**, the second and fourth order terms of the Landau expansion, and the exchange anisotropy. Eq. (1) can be simplified by rescaling the spatial variables as

$$\widetilde{w}(\mathbf{m}) = \sum_{i=1}^{3}\left(\frac{\partial \mathbf{m}}{\partial r_i}\right)^2 + 2\mathbf{m}\cdot(\nabla\times\mathbf{m}) - 2\mathbf{b}\cdot\mathbf{m} + t\mathbf{m}^2 + \mathbf{m}^4 + \sum_{i=1}^{3}\tilde{A}_e\left(\frac{\partial m_i}{\partial r_i}\right)^2, \quad (2)$$

where $\widetilde{w}(\mathbf{m}) = \frac{\beta}{K^2}w(\mathbf{M}), \mathbf{r} = \frac{\mathbf{x}}{L_D}, \mathbf{b} = \frac{\mathbf{B}}{B_0}, \mathbf{m} = \frac{\mathbf{M}}{M_0}, L_D = \frac{2A}{D}, B_0 = 2KM_0, M_0 = \sqrt{\frac{K}{\beta}}, K = \frac{D^2}{4A}, t = \frac{\alpha(T-T_0)}{K}$, and $\tilde{A}_e = \frac{A_e}{A}$. Without loss of generality, we assume that $\mathbf{b} = [0\ \ 0\ \ b]^T$, the possible phases that are considered during the temperature-magnetic phase diagram calculation are: the skyrmion lattice phase, the conical phase, and the helical phase. Within the $n$th order Fourier representation, the magnetization in the skyrmion lattice phase is described by[29]

$$\mathbf{m}_{Fn} = \mathbf{m}_0 + \sum_{i=1}^{n}\sum_{j=1}^{n_i}\mathbf{m}_{\mathbf{q}_{ij}}e^{i\mathbf{q}_{ij}\cdot\mathbf{r}}, \quad (3)$$

where $\mathbf{m}_0 = [0\ \ 0\ \ m_0]^T, |\mathbf{q}_{i1}| = |\mathbf{q}_{i2}| = |\mathbf{q}_{i3}| = \cdots = |\mathbf{q}_{in_i}| = s_iq$, $|\mathbf{m}_{\mathbf{q}_{i1}}| = |\mathbf{m}_{\mathbf{q}_{i2}}| = |\mathbf{m}_{\mathbf{q}_{i3}}| = \cdots = |\mathbf{m}_{\mathbf{q}_{in_i}}|$, $|\mathbf{q}_{1j}| < |\mathbf{q}_{2j}| < |\mathbf{q}_{3j}| < \cdots < |\mathbf{q}_{nj}|$, and $n_i$ denotes the number of reciprocal vectors whose modulus equals to $s_iq$. Here $\mathbf{m}_{\mathbf{q}_{ij}}$ can be expanded as a linear combination of three unit eigenvectors

$$\mathbf{m}_{\mathbf{q}_{ij}} = c_{i1}\mathbf{P}_{ij1} + c_{i2}\mathbf{P}_{ij2} + c_{i3}\mathbf{P}_{ij3}, \quad (4)$$

where

$$\mathbf{P}_{ij1} = \frac{1}{\sqrt{2}s_iq}[-iq_{ijy}, iq_{ijx}, s_iq]^T, \mathbf{P}_{ij2} = \frac{1}{s_iq}[q_{ijx}, q_{ijy}, 0]^T, \mathbf{P}_{ij3} = \frac{1}{\sqrt{2}s_iq}[iq_{ijy}, -iq_{ijx}, s_iq]^T \quad (5)$$

and $\mathbf{q}_{ij} = [q_{ijx}\ \ q_{ijy}\ \ 0]^T$. One immediately notices that the triple-Q representation[17] $\mathbf{m}_{triple-Q}$ of the skyrmion lattice can be obtained from eq. (3) by setting $n = 1$, $c_{12} = c_{13} = 0$, and restricting $c_{11}$ to be real. The equilibrium magnetization in the skyrmion lattice phase is obtained by minimizing $\tilde{J}(\mathbf{m}_{Fn}) = \iint \widetilde{w}(\mathbf{m}_{Fn})dS$ with respect to $c_{i1}, c_{i2}, c_{i3}, (i = 1,2,3,\ldots,n)$, $q$ and $m_0$.

In the conical phase, the magnetization reads

$$\mathbf{m}_{conical} = \begin{bmatrix} m_q\cos(\mathbf{q}\cdot\mathbf{r}) \\ m_q\sin(\mathbf{q}\cdot\mathbf{r}) \\ m_3 \end{bmatrix}, \quad (6)$$

where $\mathbf{q} = q[0,0,1]^T$. The equilibrium magnetization in the conical phase is obtained by minimizing $\tilde{J}(\mathbf{m}_{conical})$ with respect to $m_q$, $m_3$, and $q$. When $m_q$ is zero, the conical phase reduces to the ferromagnetic phase. For negative $A_c$, the direction of $\mathbf{q}$ is pinned in direction (111) in the helical phase, where the magnetization takes the form

$$\mathbf{m}_{helical} = \begin{bmatrix} \frac{\sqrt{6}}{6} & \frac{\sqrt{2}}{2} & \frac{\sqrt{3}}{3} \\ -\frac{\sqrt{6}}{3} & 0 & \frac{\sqrt{3}}{3} \\ \frac{\sqrt{6}}{6} & -\frac{\sqrt{2}}{2} & \frac{\sqrt{3}}{3} \end{bmatrix} \begin{bmatrix} m_q \cos\left(\frac{\sqrt{3}}{3} q(r_1 + r_2 + r_3)\right) \\ m_q \sin\left(\frac{\sqrt{3}}{3} q(r_1 + r_2 + r_3)\right) \\ 0 \end{bmatrix}. \tag{7}$$

The equilibrium magnetization in the conical phase is obtained by minimizing $\tilde{J}(\mathbf{m}_{helical})$ with respect to $m_q$, and $q$. At any given rescaled temperature $t$ and any given rescaled magnetic field $b$, the stable magnetic phase is determined by comparing the minimized values of $\tilde{J}(\mathbf{m}_{Fn})$, $\tilde{J}(\mathbf{m}_{conical})$, and $\tilde{J}(\mathbf{m}_{helical})$.

### III. Results and Discussion

*a) t-b Phase diagram for different values of $a_c$ when b is directed in [0, 0, 1]*

The free energy model introduced above is used to determine the effect of exchange anisotropy, represented by the coefficient $a_c$ in eq. (2), on the $t - b$ magnetic phase diagram. In Figure 1, we plot the $t - b$ phase diagram for (a) $a_c = -0.05$, (b) $a_c = -0.1$, (c) $a_c = -0.15$, and the (d) $a_c - b$ phase diagram for $t = 0$. During the calculation, the magnetization in the skyrmion lattice phase is described by $\mathbf{m}_{F8}$, which is expressed in eq. (3) by setting $n = 8$. In Figure 1(a-c), it is shown that the area of stable skyrmion phase (marked pink) expanded rapidly as $a_c$ decreases from $-0.05$ to $-0.15$. In Figure 1(d) it is shown that at $t = 0$, the skyrmion phase becomes metastable compared with the conical phase when $a_c$ is larger than $-0.064$, which illustrates that negative exchange anisotropy stabilizes the skyrmion lattice phase.

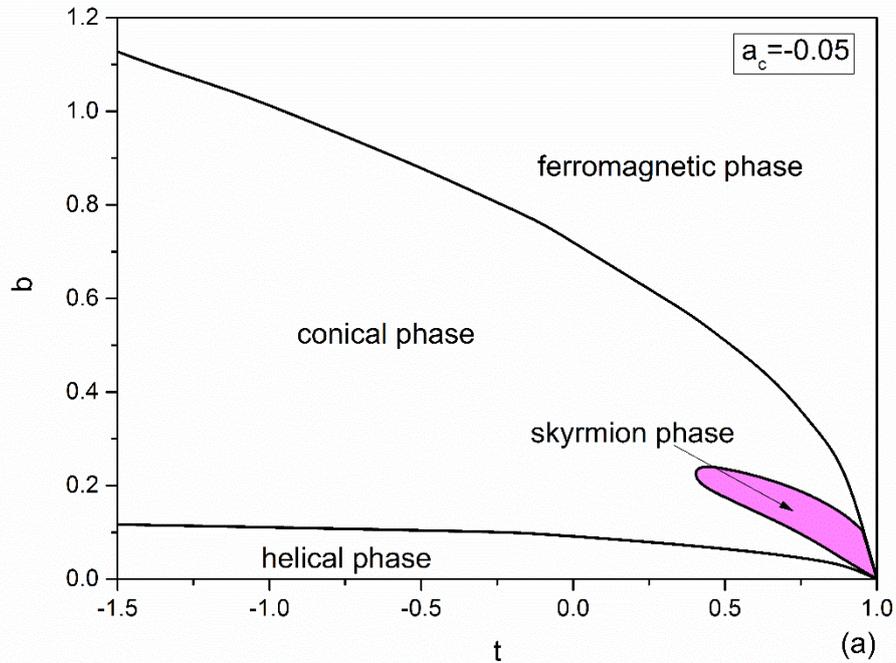

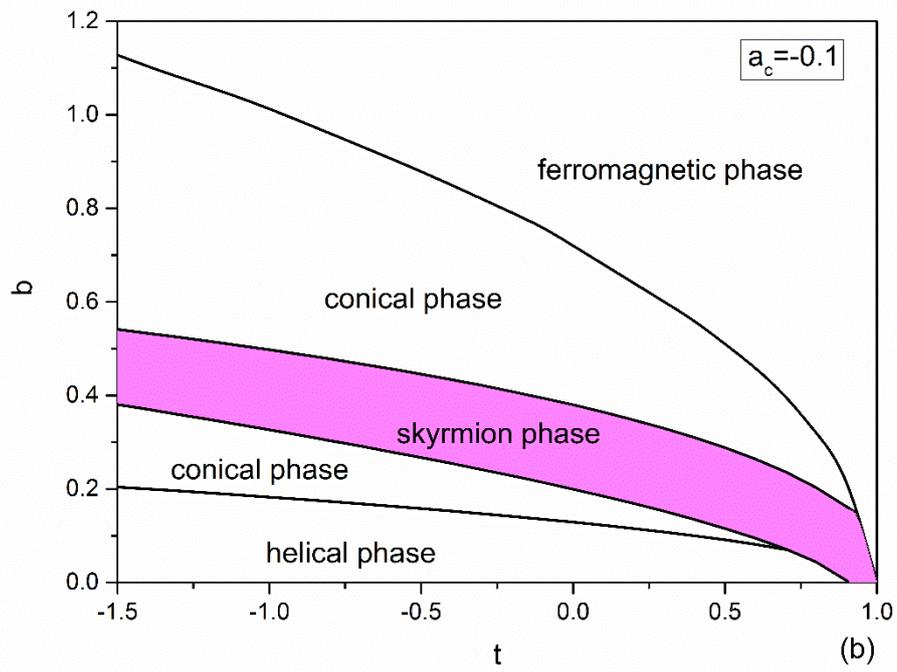

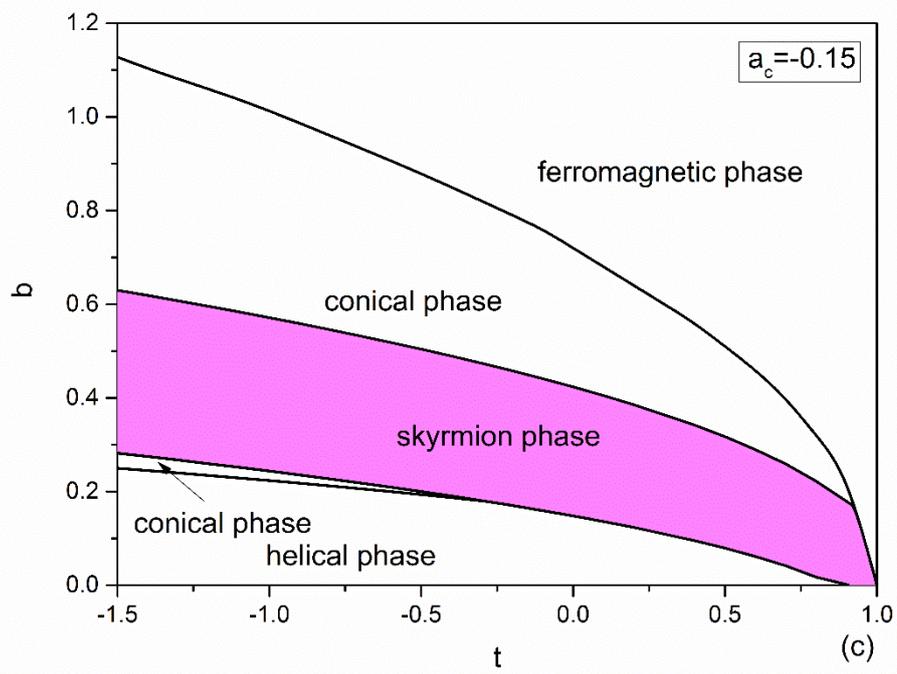

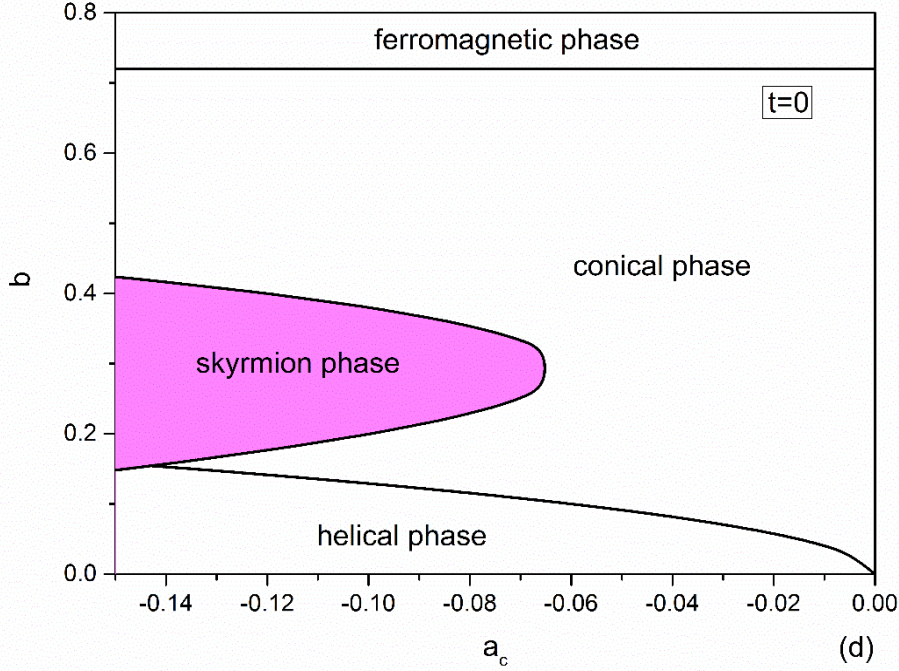

Figure 1. Magnetic phase diagrams for different strength of exchange anisotropy. $t-b$ phase diagrams for (a) $a_c = -0.05$, (b) $a_c = -0.1$, (c) $a_c = -0.15$, and (d) $a_c - b$ phase diagram for $t = 0$.

Substituting $\mathbf{m}_{conical}$ into eq. (2), one finds that the last term on the right hand side vanishes. On the other hand, in the skyrmion phase we have

$$\sum_{k=1}^{3} a_c \left(\frac{\partial (m_{Fn})_k}{\partial r_k}\right)^2 = \sum_{k=1}^{3} a_c \left[\sum_{i=1}^{n}\sum_{j=1}^{n_i} i(q_{ij})_k \left(m_{\mathbf{q}_{ij}}\right)_k e^{i\mathbf{q}_{ij}\cdot\mathbf{r}}\right]^2, \tag{8}$$

which reduces the free energy density for negative $a_c$. This explains why negative exchange anisotropy favors the skyrmion phase compared with the conical phase. Similarly, substitution of $\mathbf{m}_{helical}$ in $\sum_{i=1}^{3} a_c \left(\frac{\partial m_i}{\partial r_i}\right)^2$ yields a non-zero contribution to the free energy. Hence for the same reason, one finds that negative exchange anisotropy also favors the helical phase compared with the conical phase. For $|a_c| \ll 1$, the influence of the term $\sum_{i=1}^{3} a_c \left(\frac{\partial m_i}{\partial r_i}\right)^2$ on the equilibrium values of $m_q$ and $q$ in the helical phase is negligible. And the phase transition line between the conical phase and the helical phase in the $t-b$ phase diagram or in the $a_c - b$ phase diagram can be analytically solve as $b = \sqrt{(t-1)a_c/6}$.

### b) *Temperature-magnetic field phase diagram for MnSi using different orders of Fourier representation of the skyrmion lattice phase*

The $t-b$ phase diagram can be transformed into the temperature-magnetic field phase diagram by appropriately setting the thermodynamic parameters. Specifically, we find that by applying the thermodynamic parameters listed in Table 1 to eq. (1), we successfully reproduce the temperature-magnetic field phase diagram(Figure 1a in [17], Figure 20 in [26], and Figure 6 in [27]) for magnetic field applied along (001) quantitatively for bulk MnSi in Figure 2(a) when setting $a_c = -0.05$.

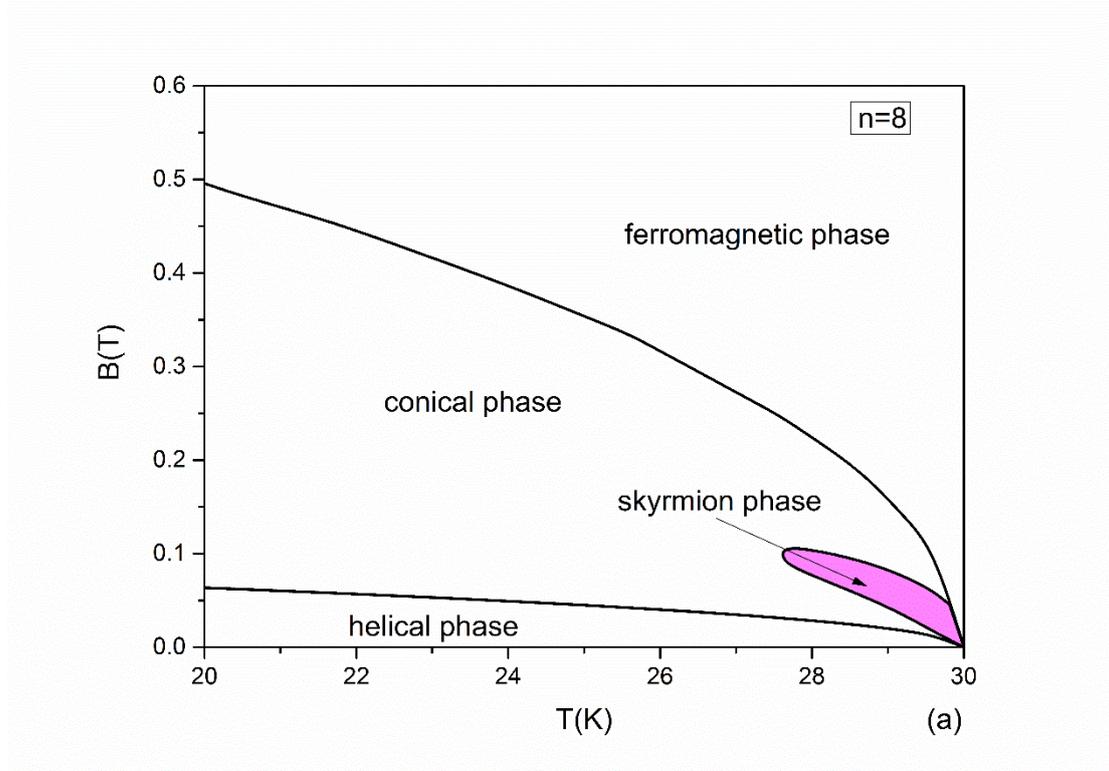

(a)

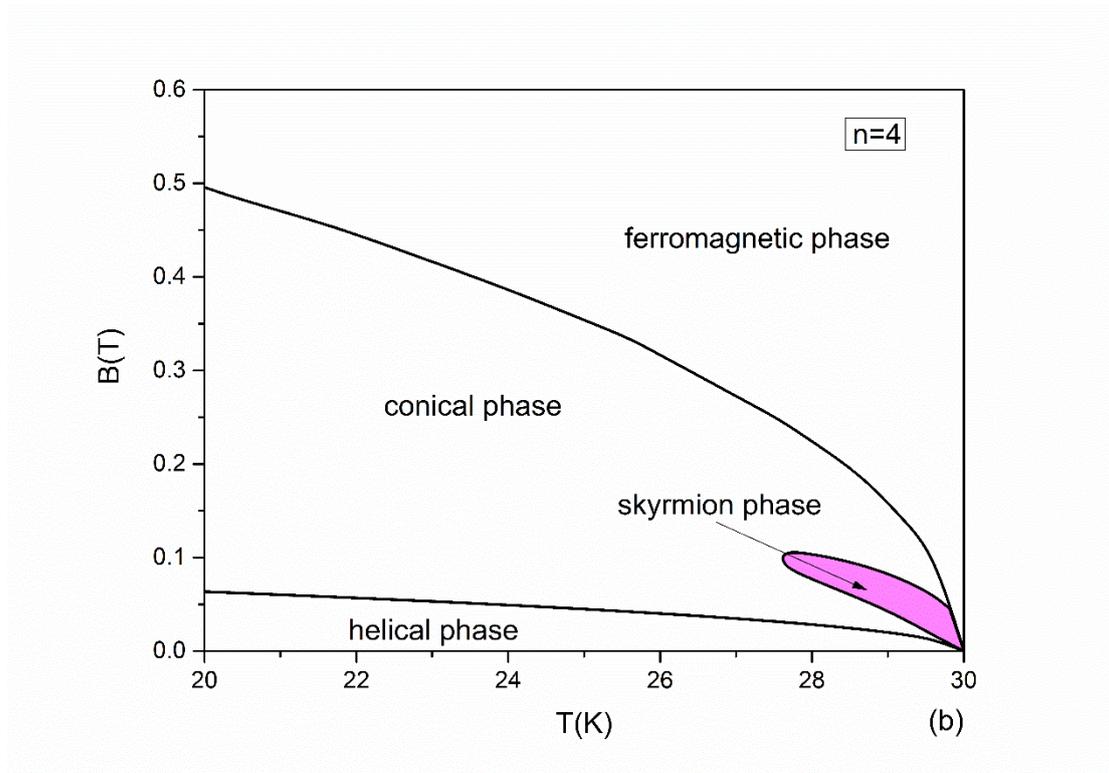

(b)

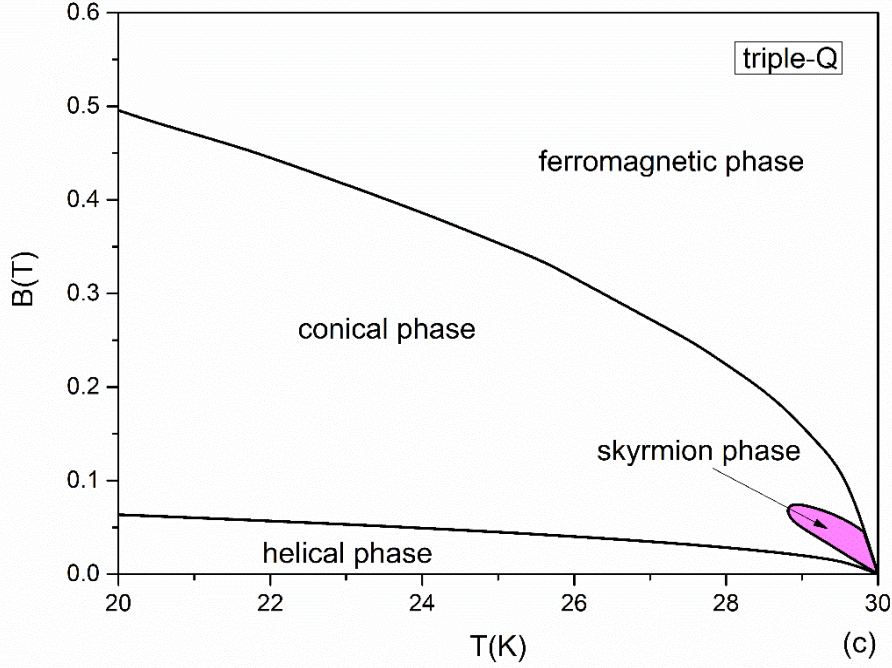

Figure 2. Temperature-magnetic field phase diagram for MnSi calculated for $a_c = -0.05$ with different orders of Fourier representation of the skyrmion lattice: (a) $n = 8$, (b) $n = 4$, and (c) the triple-Q representation.

Table 1. Thermodynamic parameters for bulk MnSi

$\alpha = 6.44 \times 10^{-7}$ JA$^{-2}$m$^{-1}$K$^{-1}$, $\beta = 3.53 \times 10^{-16}$ JA$^{-4}$m, $T_0 = 26$ K, $a_c = -0.05$

$A = 1.27 \times 10^{-23}$ JA$^{-2}$m [2, 31, 32], $D = 1.14 \times 10^{-14}$ JA$^{-2}$ [32, 33],

In practice, the Fourier representation of the skyrmion lattice has to be truncated at order $n$. An important question is how does the representation order affects the phase diagram calculation. To answer this question, we plot the temperature-magnetic field phase diagram for bulk MnSi in Figure 2(a-c), where the skyrmion lattice is described respectively by (a) $\mathbf{m}_{F8}$, (b) $\mathbf{m}_{F4}$, and (c) the triple-Q representation. It is found that the pocket-like area where the skyrmion phase is stable considerably shrinks in Figure 1(c) compared with that in Figure 1(a) and Figure 1(b). Yet since the shape of the area remains unchanged, a phase diagram calculation within the triple-Q approximation is qualitatively valid. On the other hand, Figure 1(a) and Figure 1(b) are almost indistinguishable, which indicates that 4$^{\text{th}}$ order Fourier representation of the skyrmion phase is sufficient to obtain a quantitatively valid temperature-magnetic field phase diagram.

*c) Dependence of the t-b phase diagram on the direction of applied magnetic field*

Unlike thermal fluctuation, exchange anisotropy is an intrinsic direction-dependent effect, which means that if we change the direction of applied magnetic field, we should obtain different phase diagrams through free energy minimization using eq.(2). Assume that the magnetic field is applied in direction $\left(\frac{\sqrt{2}}{2}\sin\theta, \frac{\sqrt{2}}{2}\sin\theta, \cos\theta\right)$, where $\theta$ denotes the angle between the applied field and z-axis, we perform the phase diagram calculation using eq. (2) for different values of $\theta$. For convenience, we transform the rescaled coordinates used in eq. (2) from $(r_1, r_2, r_3)$ to $(r_1', r_2', r_3')$ by $r_1' = \frac{1}{2}(1 + \cos\theta)r_1 - \frac{1}{2}(1 - \cos\theta)r_2 - \frac{\sqrt{2}}{2}\sin\theta\, r_3$ , $r_2' = -\frac{1}{2}(1 - \cos\theta)r_1 + \frac{1}{2}(1 + \cos\theta)r_2 - \frac{\sqrt{2}}{2}\sin\theta\, r_3$, and $r_3' = \frac{\sqrt{2}}{2}\sin\theta\, r_1 + \frac{\sqrt{2}}{2}\sin\theta\, r_2 + \cos\theta\, r_3$. In the new coordinates, the expressions of the magnetization for the conical phase and the skyrmion phase are given in eq. (6) and eq. (3) by replacing $r_i$ with $r_i'$, and all terms of the free energy density in eq. (2) are unchanged except the exchange anisotropy term. As shown in Figure 3, the stable area of the skyrmon lattice phase shrinks rapidly as $\theta$ increases from zero,. In Figure 3(c), we can see that at $a_c = -0.05, t = 0.9$, the skyrmion lattice phase becomes metastable for any value of $b$ when $\theta$ is larger than 22°. By comparing with the experiments of phase diagrams of MnSi[26, 27], we conclude that the skyrmion lattice phase in bulk MnSi is stabilized by a combined effect of negative exchange anisotropy and thermal fluctuation. When the magnetic field is applied in (110) or (111), the stable area of the skyrmion lattice phase considerably shrinks compared with that when the magnetic field is applied in (100), which proves the existence of negative exchange anisotropy. On the other hand, the stable area of the skyrmion lattice phase does not disappear completely when the magnetic field is applied in (110) or (111), which can only be explained by the effect of thermal fluctuation.

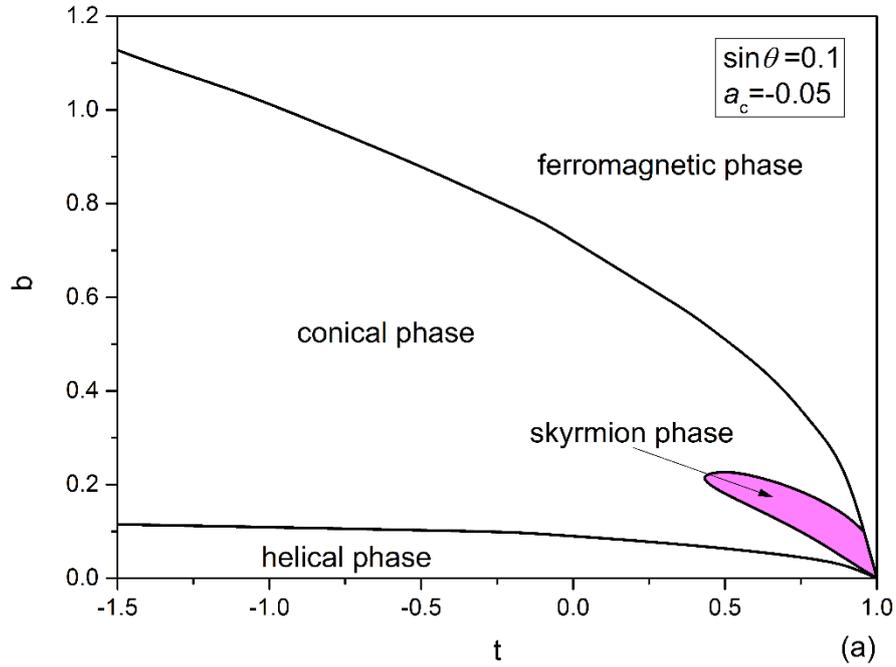

(a)

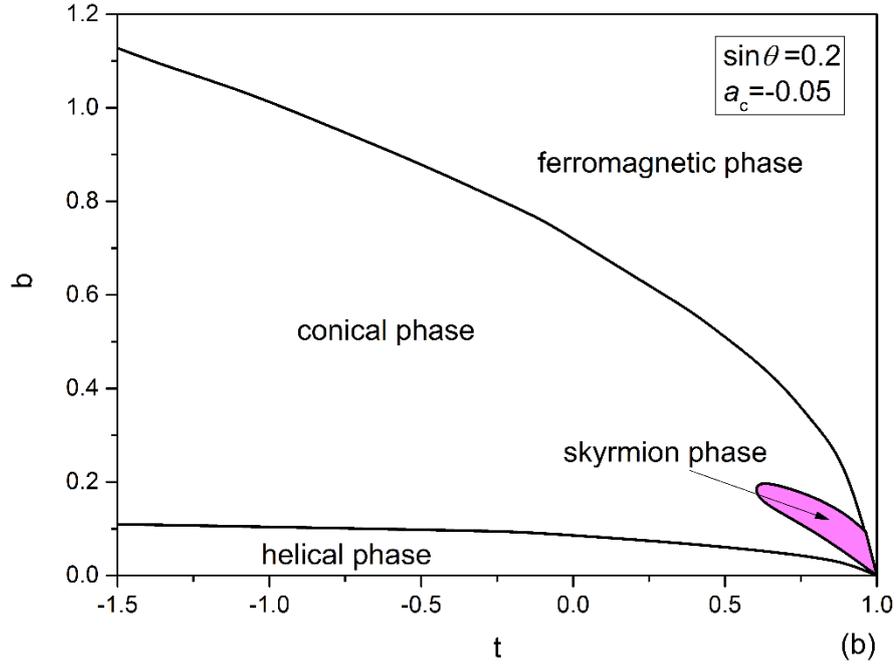

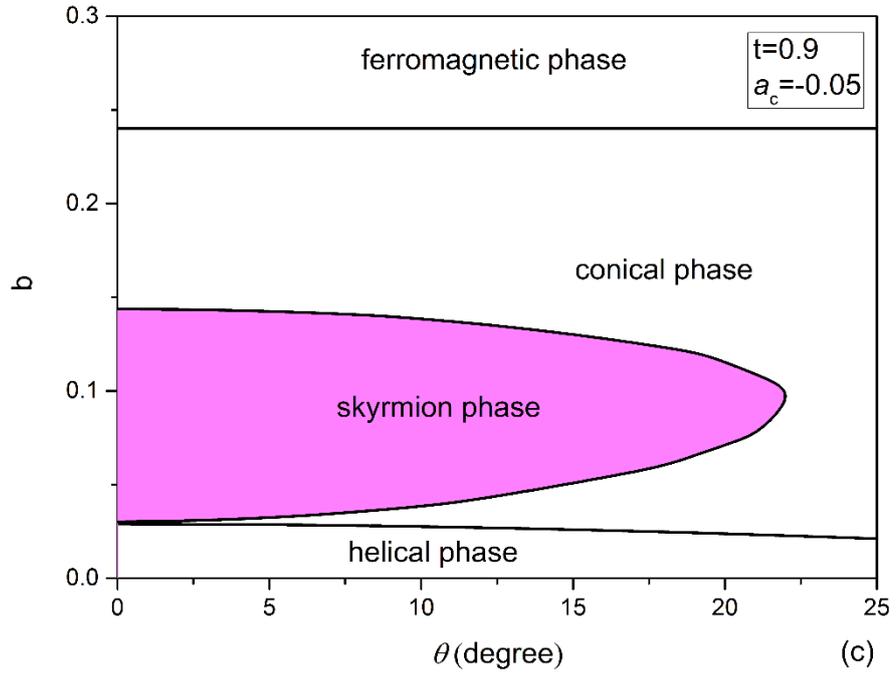

Figure 3. Magnetic phase diagrams when the magnetic field is applied in $\left(\frac{\sqrt{2}}{2}\sin\theta, \frac{\sqrt{2}}{2}\sin\theta, \cos\theta\right)$. $t-b$ phase diagrams for (a) $a_c = -0.05, \sin\theta = 0.1$, (b) $a_c = -0.05, \sin\theta = 0.2$, and $\theta - b$ phase diagram for $a_c = -0.05, t = 0.9$.

d) *Variation of the equilibrium magnetization in the skyrmion lattice phase with exchange anisotropy and Fourier representation order*

We have shown through the discussion above that negative exchange anisotropy stabilizes the skyrmion phase compared with the conical phase. A remaining question is how do the order of Fourier representation and the exchange anisotropy affect the equilibrium property of the skyrmion phase. To answer this question, distribution of $\mathbf{m}_{F8}(\mathbf{r})$ in a unit cell of the skyrmion lattice phase is plotted in Figure 4(a) at $t = 0.5$, $b = 0.2$ and $a_c = -0.05$. To clarify the effect of Fourier representation order, $[\mathbf{m}_{F8}(\mathbf{r}) - \mathbf{m}_{F4}(\mathbf{r})]/|\mathbf{m}_{F8}(\mathbf{r})|$ and $[\mathbf{m}_{F8}(\mathbf{r}) - \mathbf{m}_{triple-Q}(\mathbf{r})]/|\mathbf{m}_{F8}(\mathbf{r})|$ are plotted in Figure 4(b) and Figure 4(c) at the same condition. We see that the relative difference between $\mathbf{m}_{triple-Q}(\mathbf{r})$ and $\mathbf{m}_{F8}(\mathbf{r})$ can be as high as 7%, while the maximum relative difference between $\mathbf{m}_{F4}(\mathbf{r})$ and $\mathbf{m}_{F8}(\mathbf{r})$ reduces to less than 0.16%. The result shows that the convergence of $\mathbf{m}_{Fn}(\mathbf{r})$ with $n$ is achieved fastly. Even if we approximate the real equilibrium magnetization by the triple-Q representation, the result is qualitatively right in general. Nevertheless, if one is particularly interested in the equilibrium value of high order multi-Q components of the magnetization, high order Fourier representation is necessary. Distribution of $[\mathbf{m}_{F8}(\mathbf{r}) - \mathbf{m}_{F8}^*(\mathbf{r})]/|\mathbf{m}_{F8}(\mathbf{r})|$ in a unit cell is plotted in Figure 4(d), where $\mathbf{m}_{F8}(\mathbf{r})$ is calculated at $t = 0.5$, $b = 0.2$ and $a_c = -0.05$ and $\mathbf{m}_{F8}^*(\mathbf{r})$ is calculated at $t = 0.5$, $b = 0.2$ and $a_c = -0.1$. The result indicates that when $a_c$ decreases from $-0.05$ to $-0.1$, the relative change of the equilibrium magnetization is everywhere within 0.8%. We come to the following conclusion that while the stability area of the skyrmion phase in the phase diagram is sensitive to changes of $a_c$, the equilibrium magnetization is merely affected in the same process. This can be understood since the relative difference of free energy between the skyrmion phase and the conical phase is very small, which requires only weak exchange anisotropy to stabilize the skyrmion phase. On the other hand, the equilibrium property of the skyrmion phase is determined through competition mainly between the exchange interaction and the DM interaction. At $|a_c| = \left|\frac{A_c}{A}\right| \ll 1$, the exchange anisotropy (characterized by $A_c$) is a negligible effect compared with the isotropic exchange interaction (characterized by $A$), which is true for real material systems[2].

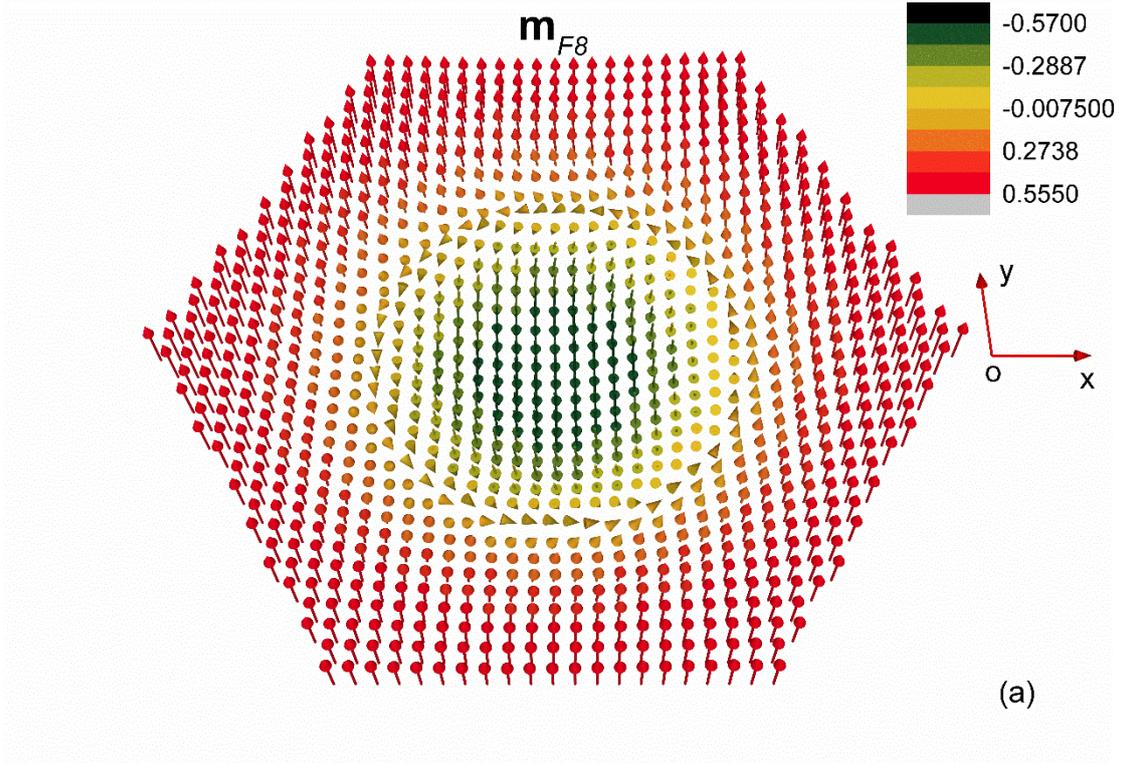

(a)

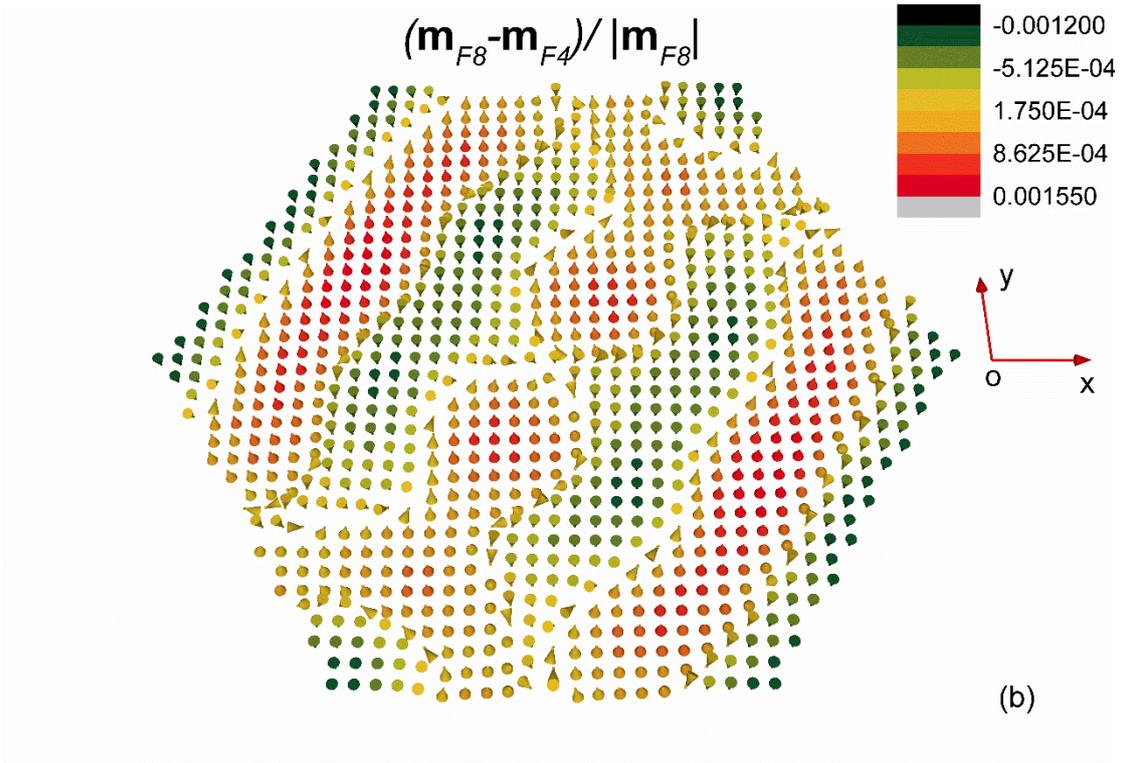

(b)

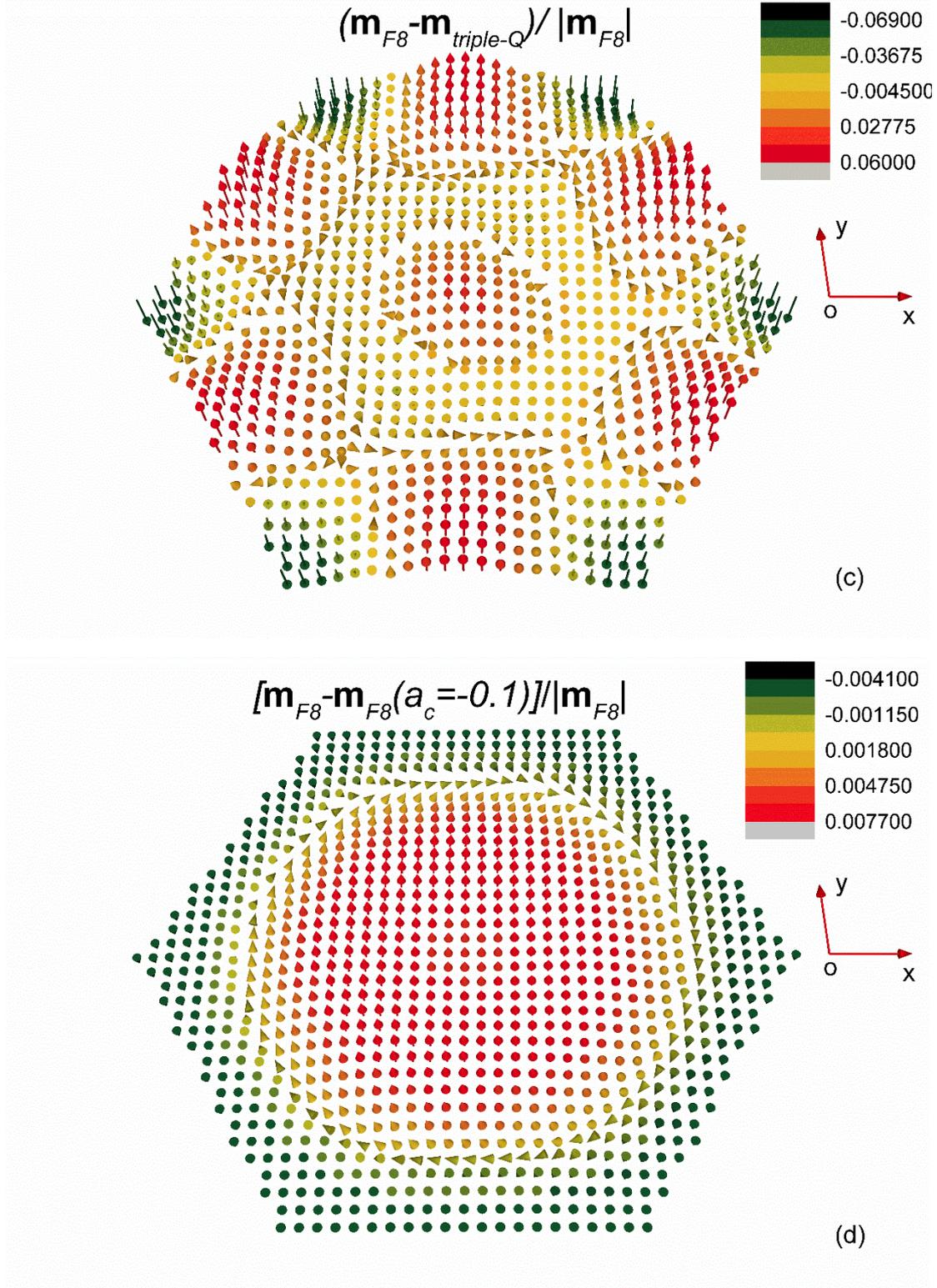

Figure 4. Distribution in a unit cell of (a) $\mathbf{m}_{F8}(\mathbf{r})$, (b) $[\mathbf{m}_{F8}(\mathbf{r}) - \mathbf{m}_{F4}(\mathbf{r})]/|\mathbf{m}_{F8}(\mathbf{r})|$, (c) $[\mathbf{m}_{F8}(\mathbf{r}) - \mathbf{m}_{triple-Q}(\mathbf{r})]/|\mathbf{m}_{F8}(\mathbf{r})|$ calculated at $t = 0.5$, $b = 0.2$ and $a_c = -0.05$. (d) Distribution of $[\mathbf{m}_{F8}(\mathbf{r}) - \mathbf{m}_{F8}^*(\mathbf{r})]/|\mathbf{m}_{F8}(\mathbf{r})|$ in a unit cell, where $\mathbf{m}_{F8}(\mathbf{r})$ is calculated at $t = 0.5$, $b = 0.2$ and $a_c = -0.05$ while $\mathbf{m}_{F8}^*(\mathbf{r})$ is calculated at $t = 0.5$, $b = 0.2$ and $a_c = -0.1$.

*e) Variation of the skyrmion lattice constant with exchange anisotropy*

Another equilibrium property of the skyrmion lattice to be discussed is the lattice constant. As mentioned in one of our previous work[29], the variation of skyrmion lattice constant to the applied magnetic field is very sensitive to the order of Fourier representation for $n \leq 4$. When using the triple-Q representation, the skyrmion lattice constant does not change with $b$ at all. Therefore we have to use at least 4$^{\text{th}}$ order Fourier representation to obtain a quantitatively valid result of the skyrmion lattice constant as well as its variation with external magnet field. We find in our calculation that the lattice constant, which is determined by $q$, as well as its variation with external magnetic field merely change with the exchange anisotropy coefficient $a_c$ in the range from 0 to $-0.2$. This can also be explained by the smallness of the exchange anisotropy compared with the isotropic exchange interaction.

## IV. Conclusion

In conclusion, we show that in bulk cubic helimagnets, the direction dependence of the stable area of the skyrmion lattice phase is the temperature-magnetic field phase diagram is caused by negative exchange anisotropy. By using the Fourier representation of the skyrmion lattice and setting $a_c = -0.05$, we quantitatively reproduce the well-known temperature-magnetic field phase diagram for bulk MnSi[17] within a phenomenological mean-field theory. In a previous work[28], exchange anisotropy is also used to explain the stability of the skyrmion lattice, where an isolated-skyrmion-based approximation is used to describe the skyrmion lattice phase. And the phase diagram obtained is slightly different from the one seen in the experiment. Our result can be considered as a support to the wave-nature of the skyrmion lattice[29].

**Acknowledgement**: The author would like to thank Achim Rosch for helpful discussion. The work was supported by the NSFC (National Natural Science Foundation of China) through the fund 11772360, 11302267, 11472313, 11572355.